\documentclass[conference]{IEEEtran}

\usepackage[utf8]{inputenc}
\usepackage[cmex10]{amsmath}
\usepackage{amsfonts}
\usepackage{amssymb}
\usepackage{graphicx}
\usepackage{float}
\graphicspath{ {./graphics/} }

\usepackage{url}

\begin{document}
\title{Human Mobility and Predictability enriched by Social Phenomena Information}

\author{
\IEEEauthorblockN{Nicolas B. Ponieman}
\IEEEauthorblockA{Grandata Labs, Argentina \\
\texttt{ nico@grandata.com } }

\and

\IEEEauthorblockN{Alejo Salles}
\IEEEauthorblockA{Physics Dept., UBA, Argentina\\
\texttt{ alejo@df.uba.ar } }

\and

\IEEEauthorblockN{Carlos Sarraute}
\IEEEauthorblockA{Grandata Labs, Argentina \\
\texttt{ charles@grandata.com } }

}

\maketitle

\begin{abstract}

The massive amounts of geolocation data collected from mobile phone records has sparked an ongoing effort to understand and predict the mobility patterns of human beings.
In this work, we study the extent to which social phenomena are reflected in mobile phone data, 
focusing in particular in the cases of urban commute and major sports events. 
We illustrate how these events are reflected in the data, and show how information 
about the events can be used to improve predictability in a simple model for a mobile 
phone user's location. 

\end{abstract}

\section{Introduction}

Mobile phone operators have access to an unprecedented volume of information
about users' real-world activities. The records of calls and messages
exchanged between their users provides a deep insight into
the interactions and activities of millions of individuals.
The social graph induced by mobile communications has provided a rich
field to apply social network analysis to real-world problems.
For instance, we can highlight the use of community 
detection techniques (see \cite{newman2004finding}\cite{blondel2008fast}\cite{fortunato2010community}\cite{wang2011community});
and the more recent advances in detecting the evolution of communities
in dynamic networks (taking into account the evolution of the social
graph over time) in \cite{aynaud2010static}\cite{sarraute2013evolution}.

A key aspect of the data collected by mobile phone operators
that has attracted considerable attention in recent years
is the information about how people are moving in the real world.
In fact, mobile phone records can be considered
as the most detailed information on human mobility across a large part
of the population \cite{song2010limits}.
The study of the dynamics of human mobility using the collected
geolocations of users, and applying it to predict future users' locations,
has been an active field of research \cite{domenico2012interdependence}\cite{lu2012predictability}.
In particular, this information
can be used to validate human mobility models (as the authors of \cite{nguyen2012using}
did with the information from a location-based social networking site);
and to study the interplay between individual mobility and social networks \cite{wang2011human}.

The study of human mobility can be applied to domains as diverse as
city planning and traffic engineering (e.g. to optimize the public transportation system
and the roads network);
public health (e.g. to allow health officials to track and predict the spread of contagious diseases);
or to guide humanitarian relief after a large-scale disaster
(see \cite{lu2012predictability} wherein the authors study population movements
after the Haiti 2010 earthquake).
Using real-world data to understand human mobility is critical to such applications.
On the business side of applications, mobile carriers are seeking for new revenue streams 
based on the anonymized and aggregated analysis of their 
subscribers' mobility data \cite{leber2013wireless}.

In this work, we study the extent to which social phenomena are reflected in mobile phone data, 
focusing in particular in the cases of urban commute and major sports events. 
The rest of the paper is organized as follows.
Section~\ref{sec:data-source} briefly describes the real-world data source
that we used for our experiments.
In Section~\ref{sec:basic-model}, we present a simple model
to predict the location of a mobile phone user, that we used as baseline of predictability.
In Section~\ref{sec:mobility-patterns}, we illustrate how urban commute can be observed in the
data, and compute basic metrics.
We also show the mobility pattern associated with a sports event (namely a soccer match).
In Section~\ref{sec:improving}, we show how information about social events can be used to
improve the predictability of the simple model.
We illustrate this idea in the case of soccer matches, and use the information
of the soccer fixture to improve location predictions.
Section~\ref{sec:conclusion} concludes the paper, and discusses ideas
for future work.

\section{Mobile Data Source} \label{sec:data-source}

\begin{figure}[H]
	\centering
	\includegraphics[width=0.5\textwidth]{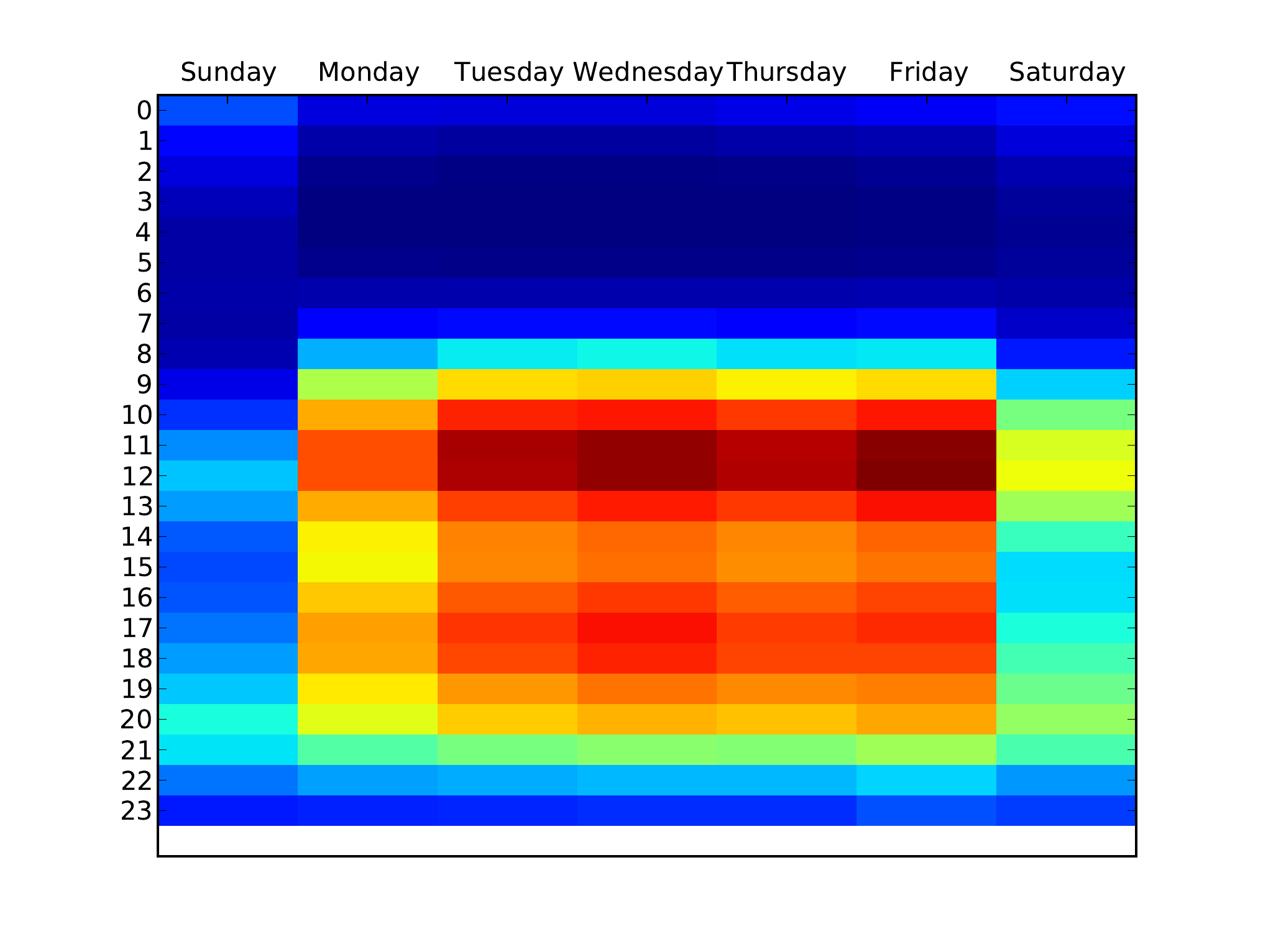}
	\caption{Call distribution according to the day of the week and the hour, 
	averaged over a period of five months, in Argentina.
	The contrast between the weekend and workweek is evident.
	It is also interesting to observe the communications peaks during the morning 
	and the afternoon from Monday through Friday. Most public holidays in the period studied where on Mondays, which is clearly visible in the figure. }
	\label{fig:call_distribution}
\end{figure}

Our data source is anonymized traffic information from a mobile operator in Argentina, focusing mostly in the Buenos Aires metropolitan area, over a period of 5 months. 
The raw data logs contain around 50 million calls per day.
Call Detail Records (CDR) are an attractive source of location information since they are collected 
for all active cellular users (about 40 million users in Argentina),
and creating additional uses of CDR data incur little marginal cost.

For our purposes, each record is represented as a tuple $\left < x, y, t, d, l \right >$,
where user $x$ is the caller, user $y$ is the callee, $t$ is the date and time of the call,
$d$ is the direction of the call (incoming or outgoing, with respect to the mobile operator client),
and $l$ is the location of the tower that routed the communication.
The temporal granularity used in this study is the hour, justified by the findings
in \cite{song2010limits}\cite{song2010modelling}.

From the operator's data, it is possible to have direct information on mobile phone usage patterns, 
as can be seen in Figure~\ref{fig:call_distribution}, which shows
the volume of communications according to the day of the week and the hour.
The expected contrast between weekend and workweek is evident. 
More interesting information is given by the communications peaks during the morning
(around 11 a.m.) and the afternoon (around 18 p.m.) from  Monday through Friday, 
which depend on the working habits in Argentina. Most of public holidays in the period studied where on Mondays, and this fact shows perfectly well in Figure~\ref{fig:call_distribution}, if we assume during holidays people show a similar calling pattern as during weekends.

\section{Mobility Model} \label{sec:basic-model}

\begin{figure}[ht]
	\centering
	\includegraphics[width=0.5\textwidth]{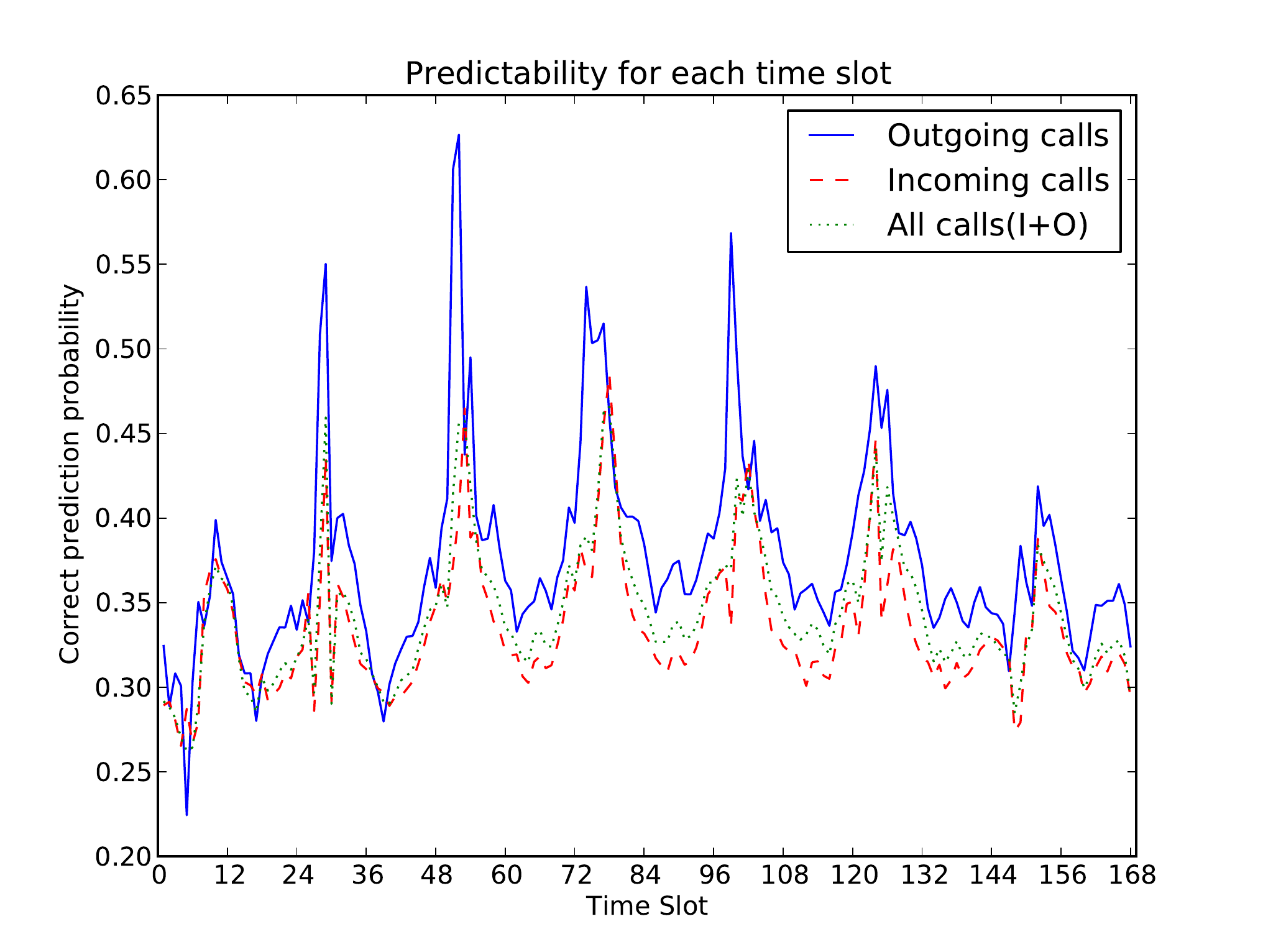}
	\caption{Users' location predictability by time slot. 
	Blue: Outgoing calls. Red: Incoming calls. Green: All calls.
	We considered the week starting at Sunday, so the first time slot corresponds to Sunday 
from midnight to 1 a.m., whereas time slot 168 corresponds to Saturday from 11 p.m to midnight.
}
	\label{fig:predictability}
\end{figure}

To predict a user's position, we use a simple model based on previous most frequent locations. We compute the correct prediction probability (i.e. accuracy) as the ratio between the number of correct predictions and total predictions made.
In order to compute these locations, we split the week in time slots, one for each hour, totalizing $7*24 = 168$ slots per week. 
Since humans tend to have very predictable mobility 
patterns \cite{song2010limits}\cite{gonzalez2008understanding}\cite{jiang2012clustering}, 
this simple model turns out to give a good predictability baseline, achieving an average of around $35 \%$ correct predictions for a period of 2 weeks, training with 15 weeks of data, 
including peaks of over $50 \%$ predictability. 
This model was used as a baseline in~\cite{cho2011friendship}, with which our results agree. 
In Figure~\ref{fig:predictability} we show the average predictability for all time slots
(considering the week from Sunday to Saturday).

Although the kind of periodic behaviour observed in the figure is widely explained in the literature,
it is important to make a few remarks about the results obtained:
\begin{itemize}
\item It is clear that predictability is at least $25\%$ higher during weekdays (Monday - Friday) than during the weekend.
\item During the night, people have a peak of predictability, corresponding to the time 
they typically spend at home.
\item Predictability is slightly higher when computed from outgoing calls than when computed from incoming calls.
\item We expect correct prediction probability to improve, and its curve to be smoother, if we filter users and make predictions only among the ones with high number of communications. This analysis will be performed in the near future.
\end{itemize}
 
It is important to notice that it is possible to improve the accuracy of this simple model by clustering antennas as
we are considering each antenna as a different location. Although this seems to be a reasonable choice, real life situations do not adjust perfectly to this schema. While a user is at her house, she might be using more than one antenna, and we are considering that she is in two different places. On the other hand, a user might use the same antenna while she is in different locations, like her workplace and the nearby restaurant where she eats lunch.
Some of these problems are pathological, and can not be tackled due to the poor space resolution given by antennas (as contrasted, for example, with GPS information). However, several problems can be solved by clustering antennas, where those clusters would represent real locations for users.

\section{Mobility Patterns} \label{sec:mobility-patterns}

\subsection{Urban Commute} \label{sec:urban-commute}

\begin{figure}[H]
	\centering
	\includegraphics[width=0.48\textwidth]{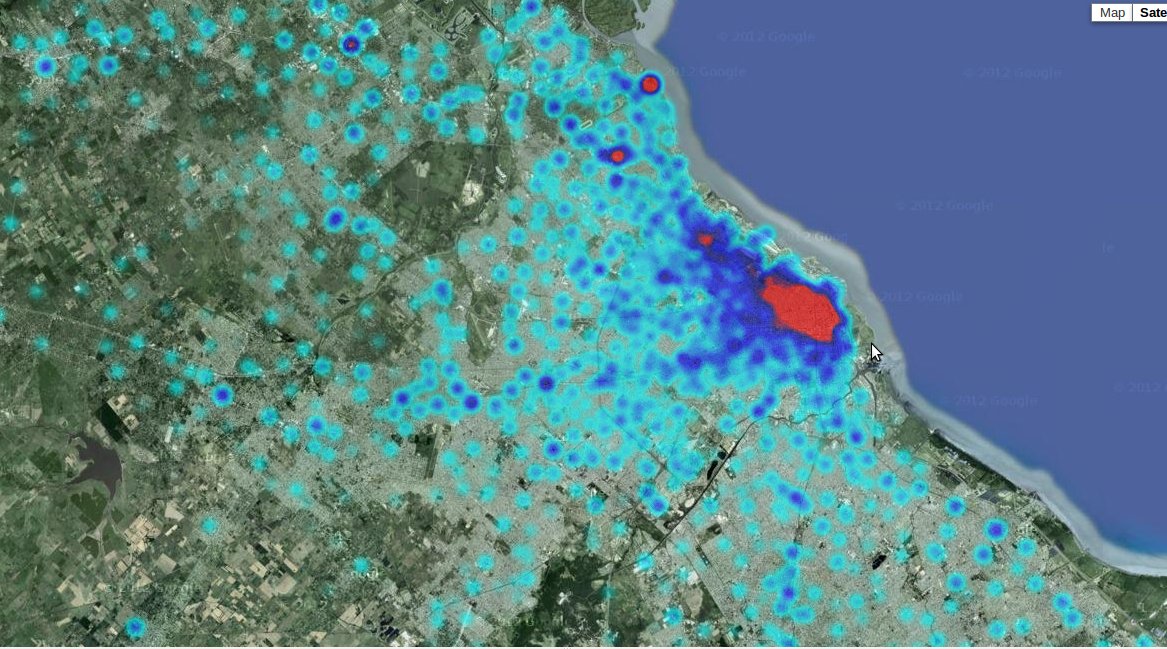}
	\caption{Antenna call distribution in Buenos Aires city and its surroundings (the Greater Buenos Aires). Note that the color scale is different than the one used in Figure~\ref{fig:commute}.}
	\label{fig:antenna_call_distribution}
\end{figure}

\begin{figure*}[th]
\begin{minipage}{.33\textwidth}
  \centering
  \includegraphics[width=0.95\textwidth]{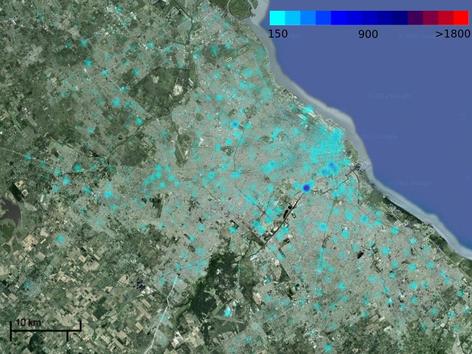}
  (a) 6 a.m.
\end{minipage}
\vspace{0.2cm}
\begin{minipage}{.33\textwidth}
  \centering
  \includegraphics[width=0.95\textwidth]{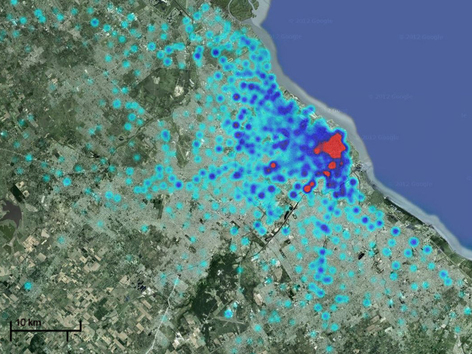}
  (b) 8 a.m.
\end{minipage}
\begin{minipage}{.33\textwidth}
  \centering
  \includegraphics[width=0.95\textwidth]{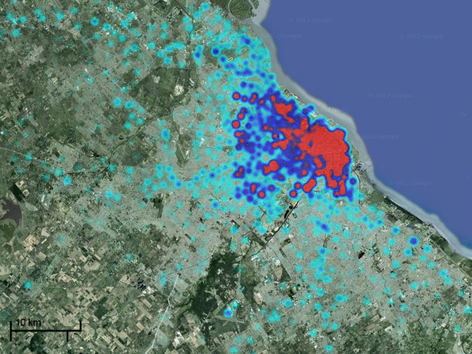}
  (c) 10 a.m.
\end{minipage}
\begin{minipage}{.33\textwidth}
  \centering
  \includegraphics[width=0.95\textwidth]{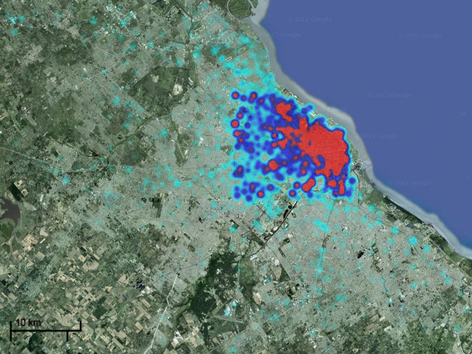}
  (d) 2 p.m.
\end{minipage}
\begin{minipage}{.33\textwidth}
  \centering
  \includegraphics[width=0.95\textwidth]{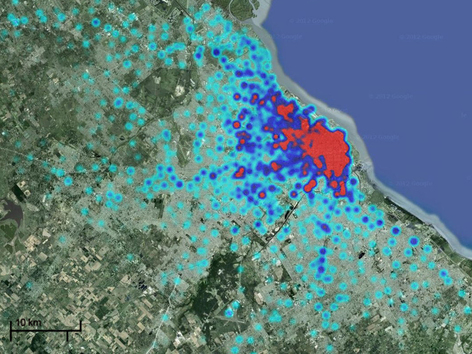}
  (e) 5 p.m.
\end{minipage}
\vspace{0.2cm}
\begin{minipage}{.33\textwidth}
  \centering
  \includegraphics[width=0.95\textwidth]{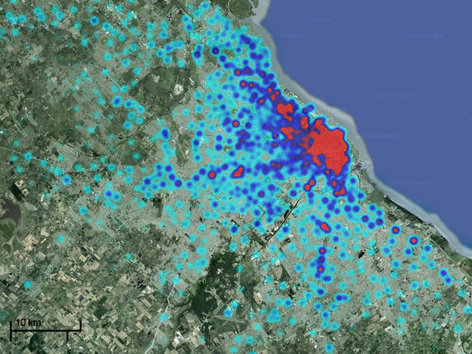}
  (f) 6 p.m.
\end{minipage}
\begin{minipage}{.33\textwidth}
  \centering
  \includegraphics[width=0.95\textwidth]{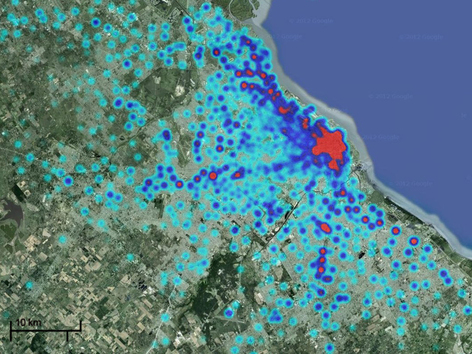}
  (g) 7 p.m.
\end{minipage}
\begin{minipage}{.33\textwidth}
  \centering
  \includegraphics[width=0.95\textwidth]{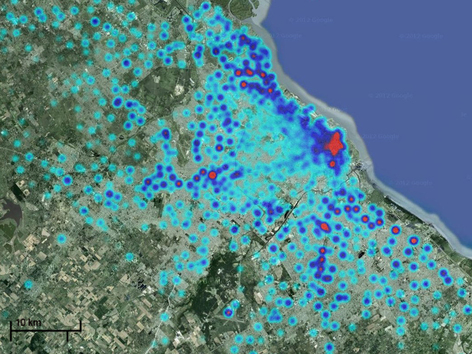}
  (h) 8 p.m.
\end{minipage}
\vspace{0.2cm}
\begin{minipage}{.33\textwidth}
  \centering
  \includegraphics[width=0.95\textwidth]{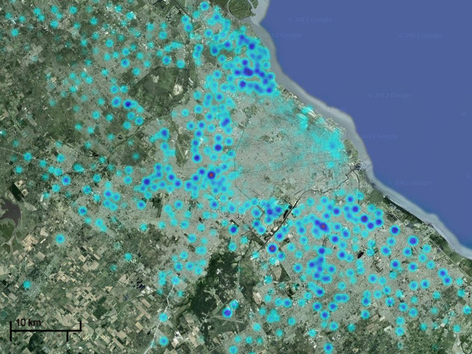}
  (i) 10 p.m.
\end{minipage}

  \caption{\label{fig:commute} Commute to Buenos Aires city from the surrounding areas on a weekday,
   for different hours.
    The color scale can be seen in image (a) where numbers represent estimated number of people in a circle from the corresponding color.
Image (g) -- corresponding to 7 p.m. -- clearly shows the major roads and highways connecting the city center to the North, West and South suburbs.
}
\vspace{-0.0cm}
\end{figure*}

The phenomenon of commuting is prevalent in large metropolitan areas (often provoking upsetting traffic jams and incidents), and naturally appears in mobile phone data.
For instance, in \cite{isaacman2011identifying} the authors study commute distances
in Los Angeles and New York areas.
Mobile data can lead to quantification of this phenomenon in terms of useful quantities, 
which are much harder to measure directly. 
Figure~\ref{fig:antenna_call_distribution} shows the call distribution for each antenna in the area of interest, averaged over a whole month.
We include a series of call patterns illustrating the Buenos Aires commute in 
Figure~\ref{fig:commute}.

\begin{figure*}[th]
\begin{minipage}{.33\textwidth}
  \centering
  \includegraphics[width=0.95\textwidth]{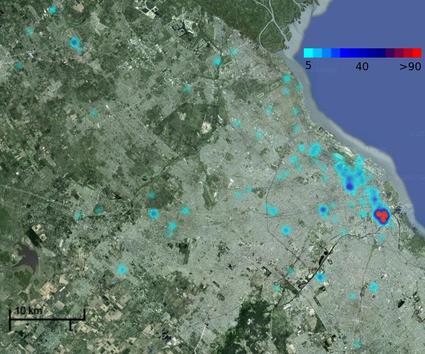}
  (a) 5 hours before
\end{minipage}
\begin{minipage}{.33\textwidth}
  \centering
  \includegraphics[width=0.95\textwidth]{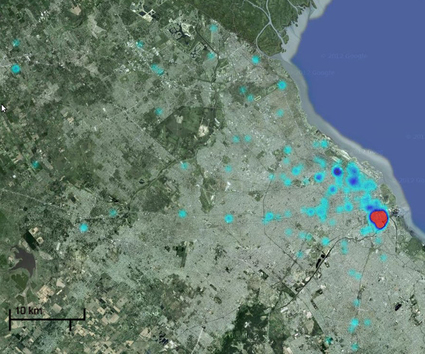}
  (b) 2 hours before
\end{minipage}
\vspace{0.2cm}
\begin{minipage}{.33\textwidth}
  \centering
  \includegraphics[width=0.95\textwidth]{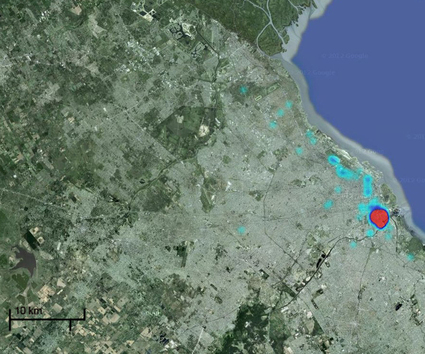}
  (c) 1 hour before
\end{minipage}
\begin{minipage}{.33\textwidth}
  \centering
  \includegraphics[width=0.95\textwidth]{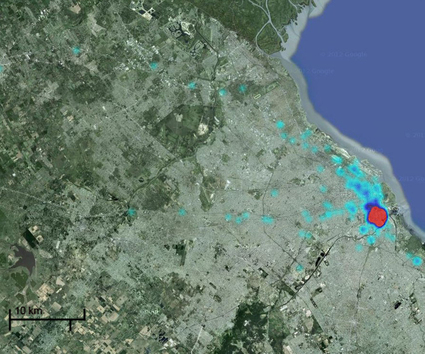}
  (d) 1 hour after
\end{minipage}
\begin{minipage}{.33\textwidth}
  \centering
  \includegraphics[width=0.95\textwidth]{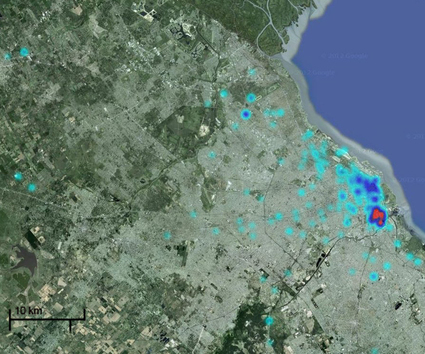}
  (e) 2 hours after
\end{minipage}
\vspace{0.2cm}
\begin{minipage}{.33\textwidth}
  \centering
  \includegraphics[width=0.95\textwidth]{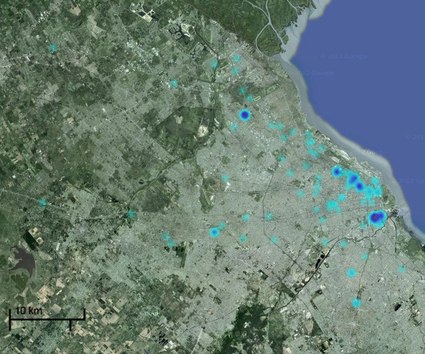}
  (f) 3 hours after
\end{minipage}
  \caption{\label{fig:boca_juega} Convergence to Boca Juniors stadium on hours prior to a soccer match, and dispersal after its end. 
        The color scale can be seen in image (a) where numbers represent estimated number of people in a circle from the corresponding color.}
\vspace{-0.0cm}
\end{figure*}
Red color corresponds to a higher number of calls, whereas blue corresponds to an intermediate number of calls and light blue to a smaller one.

\begin{figure*}[t]
\begin{minipage}{.245\textwidth}
  \centering
  \includegraphics[width=0.95\textwidth]{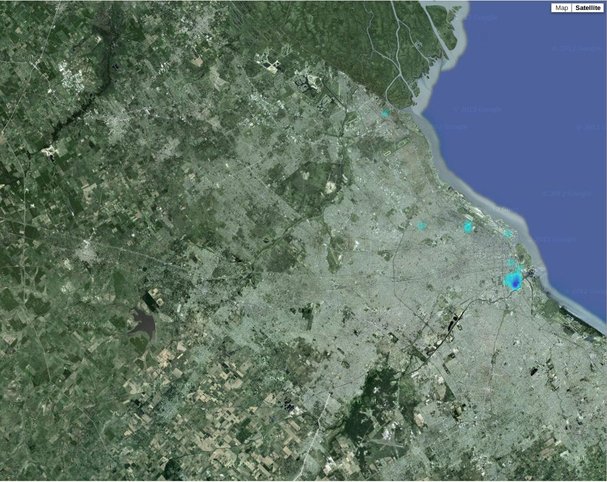}
  (a) 5 hours before
\end{minipage}
\begin{minipage}{.245\textwidth}
  \centering
  \includegraphics[width=0.95\textwidth]{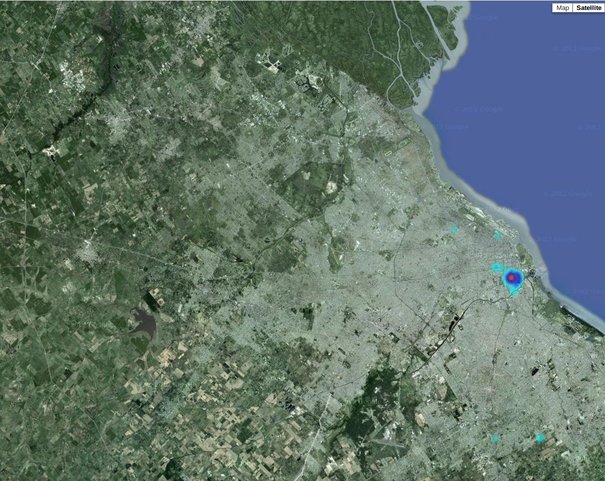}
  (b) 1 hour before
\end{minipage}
\begin{minipage}{.245\textwidth}
  \centering
  \includegraphics[width=0.95\textwidth]{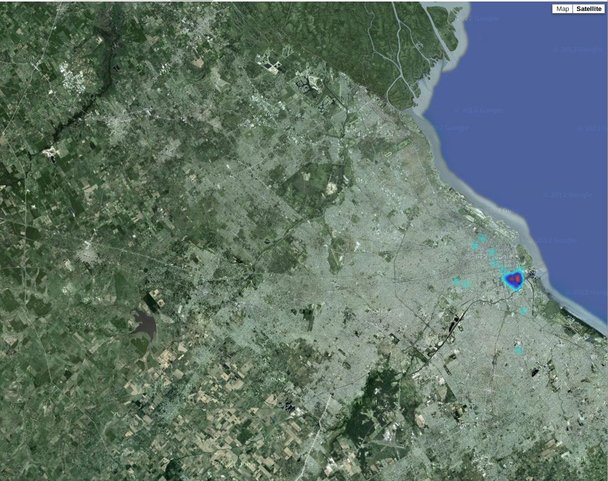}
  (c) 1 hour after
\end{minipage}
\vspace{0.2cm}
\begin{minipage}{.245\textwidth}
  \centering
  \includegraphics[width=0.95\textwidth]{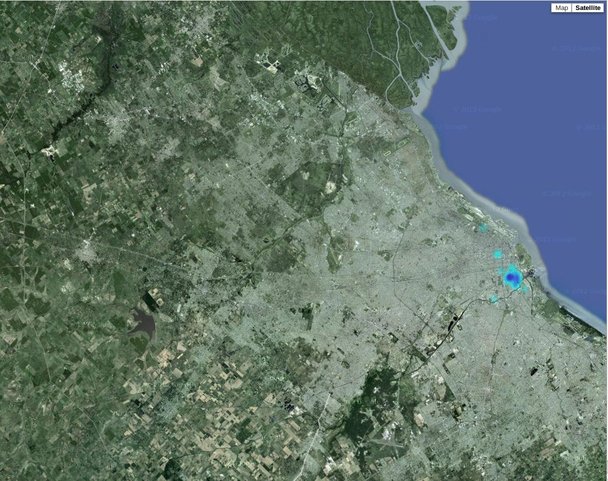}
  (d) 3 hours after
\end{minipage}
  \caption{\label{fig:no_match} Similar to figure~\ref{fig:boca_juega} on a day with no Boca match.}
\vspace{-0.0cm}
\end{figure*}

From the data, we can estimate the radius of the commute (ROC - the average distance travelled by commuters).
To do so, we take into consideration the two most frequently used antennas as the important places for each user (home and work, see \cite{csaji2012exploring}).
We proceed first to define a night time, where users are usually at their houses (9 p.m. - 5 a.m. during weekdays) and a day time where users are usually at their work places (12 p.m. - 4 p.m. during weekdays).
Afterwards, we count how many times each user makes calls in each of those two time spans both from inside and outside Buenos Aires city.

To simplify computations, we take a square area roughly corresponding to the country capital 
(the autonomous city of Buenos Aires), which is separated by a political boundary from the rest of the large metropolitan area (the Greater Buenos Aires). Around 3 million people live in Buenos Aires city, whereas around 13 million people live in the Greater Buenos Aires, which is among the top 20 largest agglomerations of the world by population.

A large part of the individuals working in Buenos Aires city live in the Greater Buenos Aires,
and commute every day. Surveys and estimations state that more than 3 million people commute to Buenos Aires city every day.
Thereby, we define a user as a commuter if she makes most of her night time calls from outside the city and most of her day calls from inside the city. 
To perform the experiment, we chose a threshold $\tau = 80\%$, meaning that 
at least $\tau$ of a user's night time calls must be made from outside the city 
and at least $\tau$ of day time calls must be made from inside the city in order 
to be considered a commuter.

After defining commuters, their home and work locations have to be found in order to compute the radius of commute. We define their home to be the antenna with the highest number of communications from outside the city during night time, and analogously define their work to be the antenna with highest number of communications inside the city during day time. We consequently consider the distance between those two main antennas as the ROC for each user. 

Having made the preceding definitions and assumptions, we compute an average ROC of $7.8$ km (as a comparison, the diameter of the city is about 14 km, and the diameter of the considered metropolitan area is 90 km).

We also computed a random ROC assuming users' locations were randomly distributed in the region of interest and, once more, defining as commuters users that live outside the city but work inside. The result was a randomized average ROC of $32.9$ km.

The previous result confirms an intuitive idea: people's living and working places tend to be closer than what a random distribution would predict.

\subsection{Sports Events} \label{sec:sports-events}

As in the urban commute case, we study human mobility in sports events as seen through mobile phone data. In Figure~\ref{fig:boca_juega}, we show how assistants to a Boca Juniors soccer match converge to the stadium in the hours prior to the game, and disperse afterwards. Average attendance to Boca Juniors home matches is 42000 people.

Note that postselecting the users attending the event necessarily produces the effect of having no calls outside the chosen area during the match. However, the convergence pattern observed is markedly different from the one seen for the same time slot of the week on a day with no match, as shown in Figure~\ref{fig:no_match}.

\section{Improving Predictability with External Data} \label{sec:improving}

So far, our results allow us to understand (and quantify) social events through the analysis of mobile phone data. This understanding can be in turn used to improve the mobility model. Social relations among individuals have been used to improve predictability in mobility models before, as in~\cite{cho2011friendship}, where social links learned from the mobile phone records are used to this end. Here, instead of peer to peer links learned from the mobile data, we show how an external data source can be used to improve the model.

We illustrate this effect using as proof of concept the case study of soccer matches.
By taking the soccer fixture, we tag users as ``Boca Juniors fans'' if they make calls using antennas around the stadium and during the time slots of Boca matches for three selected consecutive matches (which include both home and away matches), which can be considered as part of our training set for the new approach.
Using this tagging, we can dramatically improve predictability for this group of Boca fans, even predicting locations that had never been visited by a user before, 1000 km away from her usual location.

In the basic model, we predicted a user's location in a particular time slot to be her most frequent location in that particular time slot in the training set, whereas in this enriched model, we predict the stadium location (as a cluster of the antennas surrounding it) in case the user is a Boca Juniors fan and we are making predictions on a day where Boca plays a match on that stadium.
To evaluate the basic model, we use 15 weeks of data for training purposes, as described in section \ref{sec:basic-model}. For the enriched model, we use the same training data, adding the previously mentioned social information (i.e. tagging users) on three consecutive Boca matches in that period. The evaluation is made on the same testing data set in both cases, consisting on the three days where Boca plays the next matches.

The predictability of the model for these tagged users considering the fixture data rises for the days where there is a Boca Juniors match to $38\%$ -- which doubles the $19\%$ accuracy achieved by our previous model for the same set.
Moreover, the initial model is only able to make predictions in $63\%$ of events in the given set (as a consequence of a lack of information from the training set data), whereas the socially enriched model tries to predict $100\%$ of the events during match days, which make the previous results even more significant.

In order to understand these results, we illustrate with a few examples where the enriched model outperforms the simple model:
\begin{itemize}
	\item The simple model would rarely predict a user's location on a different city or in an unvisited location, whereas the enriched model would do so if the user is a Boca fan, and Boca has an away match in that city.
	\item If Boca usually plays matches on Sundays at 7 p.m. a Boca fan could have a stadium antenna as its most frequent antenna for that particular time slot. Consequently, the simple model would predict her to be in that location for any other Sunday. However, the enriched model wouldn't do so for away matches, and can even take into account that season is over, and therefore predict the location of the following most frequent antenna.
\end{itemize}

\section{Conclusion and Future Work} \label{sec:conclusion}

We illustrated how social phenomena can be studied through the lens of mobile phone data, which can be used to quantify different aspects of these phenomena with great practicality. Furthermore, we showed how including external information about these phenomena can improve the predictability of human mobility models.

Although we showed this in a specific case as a proof of concept experiment, we note that this procedure can be extended to other settings, not restricted to sports but including cultural events, vacation patterns and so on (see \cite{lu2012predictability} for a specially relevant application). The tagging obtained is useful on its own and is of great value for mobile phone operators.  
The big challenge in this line of work is to manage to include external data sources
in a systematic way.

The results obtained, as well as interesting ideas and questions related with this subject that were not addressed here, give a great perspective on future work.
A simple approach to improve these metrics is to cluster antennas in order to consider clusters as locations, which is much more real than considering a single antenna as a location.
The next step would be to keep working on more complex location prediction models, specially using social information, in order to improve the obtained accuracy and predictability rate.

Moreover, an important goal is to manage to include external data sources in a systematic way, trying to work with more and possibly bigger communities other than the ''Boca Juniors fans" community, and analyzing how social information modifies the predictions made by models during longer periods.

The communities idea shows a complementary way to use social information in order to improve 
the models' predictions, by taking tag-based predictions to the community level.
Defining, for instance, the ``Boca Juniors fans'' community, we can predict that if some users of this community make or receive calls in a certain location, other users in the community will do it as well.

\medskip

\section*{Acknowledgments}

The authors would like to thank Matias Travizano and Martin Minnoni
for their ideas and suggestions, and the anonymous reviewers for their feedback.

\medskip

\IEEEtriggeratref{7}

\bibliographystyle{unsrt}

\bibliography{mobility}

\end{document}